\documentclass[reprint,
superscriptaddress,
preprintnumbers,
nofootinbib,
 amsmath,amssymb,
 aps,
prl,
longbibliography,
]{revtex4-2}
\usepackage{CJKutf8}
\usepackage{multirow}
\usepackage{natbib}
\usepackage{mathtools}
\usepackage{xcolor}
\usepackage{graphicx}
\usepackage{dcolumn}
\usepackage{bbm}
\usepackage{hyperref}

\usepackage{feynmp-auto}

\DeclareMathAlphabet\mathbfcal{OMS}{cmsy}{b}{n}

\newcommand{\ra}{\rangle}

\newcommand{\ve}{\varepsilon}
\newcommand{\la}{\langle}

\newcommand{\eq}{\begin{equation}}
\newcommand{\eqe}{\end{equation}}
\newcommand{\eqa}{\begin{eqnarray}}
\newcommand{\eqae}{\end{eqnarray}}
\newcommand{\nn}{\nonumber}
\newcommand{\bn}{\begin{enumerate}}
\newcommand{\en}{\end{enumerate}}
\def\beq#1\eeq{\begin{align}#1\end{align}}

\def\identity{{\rlap{1} \hskip 1.6pt \hbox{1}}}
\def\iden{\identity}

\def\IZ{\mathbb{Z}}

\def\CB{{\mathcal B}}
\def\CC{{\mathcal C}}

\def\CF{{\mathcal F}}
\def\CG{{\mathcal G}}

\def\CI{{\mathcal I}}

\def\CL{{\mathcal L}}

\def\CT{{\mathcal T}}

\def\CV{{\mathcal V}}

\def\ve{\varepsilon}

\def\identity{{\rlap{1} \hskip 1.6pt \hbox{1}}}

\newcommand{\bfig}{\begin{figure}}
\newcommand{\efig}{\end{figure}}

\def\bl#1\el{\begin{align} #1 \end{align}}
\def\bg#1\eg{\begin{gather} #1 \end{gather}}

\def\bld#1\eld{\begin{aligned} #1 \end{aligned}}
\def\bgd#1\egd{\begin{gathered} #1 \end{gathered}}

\newcommand{\bra}[1]{\langle{#1}|}
\newcommand{\ket}[1]{|{#1}\rangle}

\def\topbotatom#1{\hbox{\hbox to 0pt{$#1\bot$\hss}$#1\top$}} 

\def\bubatom#1{\hbox{\hbox to 1pt{$#1($\hss}$#1)$}}

\newcommand{\RN}[1]{%
  \textup{\uppercase\expandafter{\romannumeral#1}}%
}

\newcommand{\be}{\begin{equation}}
\newcommand{\ee}{\end{equation}}
\newcommand{\ba}{\begin{align}}
\newcommand{\ea}{\end{align}}
\newcommand{\bi}{\begin{itemize}}
\newcommand{\ei}{\end{itemize}}

\let\a=\alpha \let\b=\beta \let\g=\gamma  \let\e=\epsilon
    
\let\l=\lambda \let\m=\mu \let\n=\nu    
\let\s=\sigma \let\t=\tau    
\let\w=\omega   \let\G=\Gamma \let\D=\Delta  
  \let\S=\Sigma   
 
\makeatletter
\newcommand*{\Rom}[1]{\expandafter\@slowromancap\romannumeral #1@}
\makeatother

\makeatletter
\newcommand*{\rom}[1]{\expandafter\romannumeral #1}
\makeatother

\let\nn=\nonumber

\def\beq#1\eeq{\begin{align}#1\end{align}}

\newcommand{\dbar}{
    d\kern-.20em\makebox[0pt][l]{$\bar{}$}\kern.20em
}
\newcommand{\deltabar}{
    \delta\kern-.20em\makebox[0pt][l]{$\bar{}$}\kern.20em
}

\begin{document}
\begin{CJK*}{UTF8}{mj}
\preprint{CERN-TH-2026-267}
\title{
Unitarity Methods for Magnus Amplitudes
}

\author{Jung-Wook Kim (김정욱)}
\email{jung-wook.kim@cern.ch}
\affiliation{Theoretical Physics Department, CERN, 1211 Geneva 23, Switzerland}

\author{Sungsoo Kim}
\email{sooo4017@snu.ac.kr}
\affiliation{Department of Physics and Astronomy, Seoul National University, 1 Gwanak-ro, Gwanak-gu, Seoul 08826, Korea}

\author{Jian-Rong Li}
\email{lijr07@gmail.com}
\affiliation{Faculty of Mathematics, University of Seville, Calle Tarfia s/n, 41012 Seville, Spain}

\begin{abstract}
Magnus amplitudes refer to the matrix elements of the logarithm of the $S$-matrix. Modern unitarity methods construct cut-constructible parts of scattering amplitudes without directly invoking Feynman diagrams, offering an efficient way of obtaining loop integrands for scattering amplitudes. A procedure is proposed that adapts unitarity methods to the computation of Magnus amplitudes by modifying the $i0^+$ prescriptions of scattering amplitude integrands. A sufficient condition for applying the procedure at the level of master integrals is identified.
\end{abstract}

\maketitle
\end{CJK*}

\section{Introduction}
\emph{Magnus amplitudes} refer to the matrix elements of the \emph{Magnusian} $\chi$, which is a Hermitian matrix that explicitly expresses the $S$-matrix as a unitary matrix: $S = e^{i \chi}$~\cite{Kim:2025gis,Brandhuber:2025igz}. What does it mean to apply unitarity methods to a manifestly unitary representation of the $S$-matrix?

The motivation for studying the exponential representation of the $S$-matrix comes from gravitational wave physics. 
Since the first direct detection of gravitational waves~\cite{LIGOScientific:2016aoc}, the research on the application of modern $S$-matrix techniques to the study of gravitational waves has expanded substantially and has developed into the field of post-Minkowskian dynamics~\cite{Bern:2019nnu,Bern:2019crd,Kalin:2020mvi,Kalin:2020fhe,Bern:2021dqo,Dlapa:2021npj,Jakobsen:2022psy,Kalin:2022hph,Dlapa:2022lmu,Dlapa:2024cje,Driesse:2024xad,Bern:2024adl,Driesse:2024feo,Bern:2025zno,Bern:2025wyd,Driesse:2026qiz,Bern:2026oqp,Bjerrum-Bohr:2022blt,Kosower:2022yvp,Buonanno:2022pgc}. 
Because the ultimate interest is to extract the \emph{classical} dynamics from the \emph{quantum} $S$-matrix, several methods have been developed to facilitate the classical limit~\cite{Damgaard:2019lfh,Mogull:2020sak,Brandhuber:2021eyq}. 
The exponential representation of the $S$-matrix $S = e^{i \chi}$ was motivated by the eikonal approximation, which simplifies the classical limit to computing the leading term in the $\hbar$ expansion of the $\chi$-matrix~\cite{Damgaard:2021ipf,Damgaard:2023ttc}. 
The term \emph{Magnusian} was coined to refer to the $\chi$-matrix and its generalization to non-scattering kinematics~\cite{Kim:2025gis} and led to a flurry of studies on Magnusian-based approaches to gravitation~\cite{Kim:2025hpn,Alessio:2025flu,Kim:2025olv,Haddad:2025cmw,Kim:2025sey,Gonzo:2026yha,Bautista:2026fcp,Blanco:2026prj,Jeong:2026pyw,Ochirov:2026ixx,Moynihan:2026xmv,Placidi:2026xug,DaSilva:2026pdk,Brandhuber:2026dkd}. 
An intriguing observation is that $\chi$-matrix elements have less severe infrared divergences~\cite{Kim:2024svw,Bautista:2026fcp,Brunello:2026rdk,Brunello:2026lzf}, which may also have implications for non-gravitational theories.

The term ``Magnusian'' is derived from the Magnus series~\cite{Magnus:1954zz}. 
The formal expansion of the $S$-matrix is given by the Dyson series~\cite{Dyson:1949bp}. 
Likewise, the formal expansion of the logarithm of the $S$-matrix (the Magnusian) is given by the Magnus series. 
The $S$-matrix elements (scattering amplitudes) can be diagrammatically expanded using Feynman diagrams, and their diagrammatic rules (Feynman rules) can be derived from the Dyson series. 
Likewise, the matrix elements of the Magnusian (Magnus amplitudes) can be diagrammatically expanded using \emph{Magnus diagrams}, and their diagrammatic rules can be derived from the Magnus series~\cite{Kim:2024svw,Brandhuber:2025igz,Kim:2025ebl,Guo:2026xaw,Guo:2026hyi}. 
The aim of this study is to devise a novel computational method for Magnus amplitudes that builds on modern techniques for computing scattering amplitudes.

\emph{Efficiency matters}. The modern amplitudes program is characterized by computational methods for scattering amplitudes that are more efficient than Feynman diagrams. 
One of its major tools is generalized unitarity. 
\emph{Generalized unitarity}~\cite{Bern:1994zx,Bern:1994cg,Bern:2011qt} constructs the (cut-constructible parts of the) loop integrand of scattering amplitudes by merging tree-level scattering amplitudes and exploits the simplicity of (on-shell) scattering amplitudes, which is not apparent at the level of individual Feynman diagrams. 
Returning to the question raised at the beginning, the answer is to adapt generalized unitarity to the computation of Magnus amplitudes.

\textit{Notation}: Scattering \textbf{A}mplitudes are denoted by $A$, and \textbf{M}agnus amplitudes are denoted by $M$. 
Graph isomorphism is denoted by $\cong$.
The sum over graphs $\sum$ counts isomorphic graphs only once, while the sum over graphs $\sideset{}{'}{\textstyle\sum}$ counts all instances of isomorphic graphs separately.
An overbarred graph $\bar{G}$ is an undirected graph. 
The symmetry factor $S(\bar{G})$ does not consider edge orientations, while the symmetry factor $\s (G)$ distinguishes edge orientations. 
Both symmetry factors take into account decorations (momentum inflow (for vertices), masses, cuts, propagator powers, etc.).

\section{Review of generalized unitarity}

Generalized unitarity reconstructs the loop integrand from residues at selected propagator poles~\cite{Bern:1994zx,Bern:1994cg,Bern:2011qt}. When inverse propagators $\{ D_0, D_1,\cdots, D_n \}$ are selected as cuts, the loop integrand is constrained to take the form
\begin{align}
    A \sim \int_{\{ \ell \}} \frac{N(\{ \ell \}) + \cdots}{D_0 D_1 \cdots D_n} \,, \label{eq:gen_uni_form}
\end{align}
where $\{ \ell \}$ collectively denotes loop momenta and the ellipsis denotes terms that vanish on the simultaneous cut $D_0 = \cdots = D_n = 0$.

The key idea of generalized unitarity is that the numerator $N(\{ \ell \})$ can be constructed as a sum of products of scattering amplitudes, where the sum runs over all possible polarizations and species of the particles on the cuts. For example, generalized unitarity applied to the triple cut condition $D_0 = D_1 = D_2 = 0$ can be diagrammatically represented as
\begin{align}
\begin{gathered}
    A \sim \int_{\{ \ell \}} \frac{N( \{ \ell \} )}{D_0 D_1 D_2} + \cdots \;\; \sim \quad\;
    \begin{fmffile}{triple_cut_example}
    \parbox{75pt}{
        \begin{fmfgraph*}(75,75)
            \fmfstraight
            \fmfleft{l3,l2,l1}
            \fmfright{r3,r2,r1}
            \fmf{phantom}{l2,e2,cl2,e1,cr1,e3,r2}
             \fmffreeze
            \fmf{phantom}{l1,cl1,cl2,cl3,l3}
             \fmffreeze
            \fmf{plain,width=3}{l1,cl1}
            \fmf{plain,width=3}{l3,cl3}
            \fmf{plain,width=3}{r2,cr1}
            \fmf{plain,width=2,label=$D_0$}{cl1,cl3}
            \fmf{plain,width=2,label=$D_1$,label.side=right}{cl3,cr1}
            \fmf{plain,width=2,label=$D_2$,label.side=right}{cr1,cl1}
            \fmfv{decor.shape=circle,decor.filled=empty,decor.size=25,label=$A_1$,label.dist=0}{cl1}
            \fmfv{decor.shape=circle,decor.filled=empty,decor.size=25,label=$A_2$,label.dist=0}{cl3}
            \fmfv{decor.shape=circle,decor.filled=empty,decor.size=25,label=$A_3$,label.dist=0}{cr1}
            \fmfv{label=$v_1$,label.angle=-120}{l1}
            \fmfv{label=$v_2$,label.angle=120}{l3}
            \fmfv{label=$v_3$,label.angle=90}{r2}
        \end{fmfgraph*}
    }
    \end{fmffile} \;\;
    \,, 
    \\ N ( \{ \ell \} ) \sim \sum A_1 ( \{ \ell \} ) \times A_2 ( \{ \ell \} ) \times A_3 ( \{ \ell \} ) \,,
\end{gathered}
\end{align}
where external legs represent sets of external particles and the cut propagators are represented by internal edges. The blobs correspond to scattering amplitudes (subamplitudes). The sum runs over all possible polarization states and particle species on the cuts.

In generalized unitarity, loop momenta are complexified, and cut conditions put internal legs attached to subamplitudes on shell. While on-shell conditions are not necessary for external particles, it is customary to consider external particles as on shell and consider all subamplitudes as on-shell tree amplitudes, which are generally much simpler than their off-shell counterparts.

\subsection{Generalized unitarity as diagram reorganization}

Scattering amplitudes can be considered as a sum over Feynman diagrams, which are undirected graphs.
\begin{align}
    i A = \int \sum_{\bar{G}} \frac{1}{S(\bar{G})} \CI(\bar{G}) \,. \label{eq:amp_Feyn_sum}
\end{align}
Here $\CI (\bar{G})$ denotes the \emph{integrand} corresponding to the graph $\bar{G}$ constructed as the product of vertex rules and propagators. 

Imposing the cut conditions $D_0 = D_1 = \cdots = D_n = 0$ to extract the numerator $N(\{ \ell \})$ of \eqref{eq:gen_uni_form} is equivalent to selecting the diagrams containing the edges $\{ e_0 , \cdots, e_n \}$ that respectively correspond to the propagators $\{ D_0 , \cdots, D_n \}$.
\begin{align}
    i A \supset \int_{\{ \ell \}} \frac{N(\{ \ell \})}{D_0 D_1 \cdots D_n} \supset \int \sum_{\substack{\bar{G} \\ e_0, \cdots , e_n \in \bar{G} }} \frac{1}{S(\bar{G})} \CI(\bar{G}) \,. \label{eq:gen_uni_ansatz1}
\end{align}
Cutting the edges $\{ e_0 , \cdots , e_n \}$ decomposes the original graph $\bar{G}$ into the set of connected components (subgraphs) $\{ \bar{G}_1 , \cdots, \bar{G}_k \}$. 
\begin{align}
    \bar{G} \backslash \{ e_0 , \cdots , e_n \} = \bar{G}_1 \sqcup \cdots \sqcup \bar{G}_k \,.
\end{align}
The symmetry factor $S(\bar{G})$ decomposes into the symmetry factors of the subgraphs $S(\bar{G}_j)$ and cut gluings $S(\{ e \})$,
\begin{align}
    S(\bar{G}) = S(\{ e \}) \times S(\bar{G}_1) \times \cdots \times S (\bar{G}_k) \,. \label{eq:symfac_decomp}
\end{align}
The edges correspond to propagators, which can be schematically represented as
\begin{align}
    \la \CT \phi_1 \phi_2 \ra \simeq \frac{- i \sum \varepsilon^\ast (p) \varepsilon (p) + \cdots }{D(p)} \,,
\end{align}
where the numerator is the state sum over the outer product of polarization states up to terms that vanish on shell. Therefore, the integrand $\CI(\bar{G})$ decomposes into a product of ``subintegrands'' summed over all possible polarization states of the cut propagators,
\begin{align}
    \CI(\bar{G}) = \frac{(-i)^{n+1}}{D_0 D_1 \cdots D_n} \times \sum_{\{e\} }\CI(\bar{G}_1) \times \cdots \times \CI(\bar{G}_k) \,,
\end{align}
where $\sum_{\{e\}}$ denotes the sum over intermediate states on the cuts.

Collecting these ingredients, the sum \eqref{eq:gen_uni_ansatz1} can be reorganized as
\begin{align}
\begin{aligned}
    \! iA &\supset \int \frac{1}{S(\{e\})} \times \frac{(-i)^{n+1}}{D_0 D_1 \cdots D_n}
    \\ & \phantom{=asdfasdf}\times \sum_{\{e\} } \sum_{\bar{G}_1,\cdots,\bar{G}_k} \prod_{j=1}^k \frac{1}{S(\bar{G}_j)} \CI (\bar{G}_j)
    \\ &= \int \frac{1}{S(\{e\})} \times \frac{(-i)^{n+1}}{D_0 \cdots D_n} \times \sum_{\{e\} } i A_1 \times \cdots \times i A_k \,,
\end{aligned} \label{eq:genuni_deriv}
\end{align}
where each subamplitude arises from the subgraph sum $i A_j = \sum_{\bar{G}_j} \frac{1}{S(\bar{G}_j)} \CI (\bar{G}_j)$. 

\section{Applying generalized unitarity to Magnus amplitudes}

The qualitative difference between Magnus amplitudes and scattering amplitudes is that Magnus amplitudes are given as a sum over Magnus diagrams, which are \emph{directed-undirected} graphs.
\begin{align}
    i M = \int \sum_{G} \frac{\w(G)}{\s(G)} \CI (G) \,. \label{eq:mag_amp_diag_sum}
\end{align}
The weight factor $\w (G)$ is the \emph{Murua coefficient} of the graph $G$~\cite{Murua}.
The direction of each edge denotes the causal flow of the corresponding causal (retarded or advanced) propagator, and the undirected edges denote the corresponding (normalized) Hadamard cut propagator $\frac{1}{2} \la \{ \phi_1 , \phi_2\} \ra$  (see \textsc{End Matter} for differences in normalization conventions). 
With this normalization of two-point functions, the \emph{vertex rules are the same as those of Feynman diagrams} when constructing the integrand $\CI (G)$; 
therefore, the integrand $\CI(G)$ differs from the Feynman diagram integrand $\CI(\bar{G})$ only by the choice of propagators~\cite{Guo:2026xaw}. 
If the graph $G$ contains a cycle corresponding to a closed causal loop, the integral vanishes ($\int \CI(G) = 0$) although the associated Murua coefficient $\w(G)$ need not vanish.

\subsection{Magnus amplitudes from Murua \texorpdfstring{$i0^+$}{i0+} prescription of scattering amplitudes} \label{sec:Murua_prescription}

Consider the set of Magnus diagrams $\CB(\bar{G})$ whose elements represent all possible orientation configurations on $\bar{G}$. 
The key relation is the sum rule obeyed by the Murua coefficients~\cite{Kim:2024svw,Guo:2026hyi},
\begin{align}
    \sum_{G \in \CB (\bar{G})} \frac{\w(G)}{\s(G)} = \frac{1}{S(\bar{G})} \,. \label{eq:Murua_sum_rule}
\end{align}

Consider the restriction of the sum \eqref{eq:mag_amp_diag_sum} to directed Magnus diagrams---Magnus diagrams consisting solely of causal propagators,
\begin{align}
    i M \supset \int \sum_{\bar{G}} \sum_{G \in \CB (\bar{G})} \frac{\w(G)}{\s(G)} \CI (G) \,.
\end{align}
For $G \in \CB(\bar{G})$, the integrands $\CI(G)$ differ only in their $i0^+$ prescriptions of the propagators. In other words, $\CI(G) = \CI(\bar{G})$ when neglecting $i0^+$ prescriptions, and the inner sum can be reorganized as
\begin{align}
    \sum_{G \in \CB (\bar{G})} \frac{\w(G)}{\s(G)} \CI (G) = \sum_{G \in \CB (\bar{G})} \frac{\w(G)}{\s(G)} \CI (\bar{G}) = \frac{\CI (\bar{G})}{S(\bar{G})} \,.
\end{align}
Therefore, the integrands for Magnus amplitudes can be constructed from the integrand of scattering amplitudes $\frac{\CI (\bar{G})}{S(\bar{G})}$ by modifying the $i0^+$ prescriptions.\footnote{The diagrams for matrix elements of $S = e^{i \chi}$ coming from iterations of $\chi$ must have cut propagators; if none of the propagators can go on shell, the diagram must come from $\chi$. This observation extends to diagrams with cut propagators: the diagrams coming from iterations of $\chi$ necessarily have cut propagators that decompose the diagram into connected components.} 
This will be the main strategy for applying generalized unitarity to Magnus amplitudes: first construct the scattering amplitude integrands, then modify their $i0^+$ prescriptions to compute Magnus amplitudes~\cite{Kim:2024svw}. The weight $\m (G)$ of the $i0^+$ prescription corresponding to the graph $G$ can be computed as
\begin{align}
    \m (G) = \frac{\w(G)}{\s(G)} \times S(\bar{G}) \,, \label{eq:Murua_weight_nocut}
\end{align}
which sums to unity for a given topology, i.e.,
\begin{align}
    \sum_{G \in \CB (\bar{G})} \m (G) = S(\bar{G}) \times \sum_{G \in \CB (\bar{G})} \frac{\w(G)}{\s(G)} = 1 \,.
\end{align}

Extension to Magnus diagrams with cut propagators proceeds in a similar manner. Consider a skeleton graph $\bar{\CG}$ which represents the underlying Feynman diagram. 
The calligraphic letter implies that some edges may remain undirected in $\bar{\CG}$. 
Fix the set of edges $\CC \subset \bar{\CG}$ that leaves the graph $\bar{\CG} \backslash \CC$ connected, where $\bar{\CG} \backslash \CC$ is obtained from $\bar{\CG}$ by removing the edges in $\CC$. 
The edges in $\CC$ will correspond to cut propagators in Magnus diagrams. Consider the set of Magnus diagrams $\CB(\bar{\CG} \backslash \CC) \sqcup \CC$ whose elements correspond to graphs obtained by assigning an orientation to each edge of $\bar{\CG}\backslash\CC$. 
Because $\bar{\CG} \backslash \CC$ is still a connected graph~\cite{Guo:2026hyi}, 
\begin{align}
    \sideset{}{'}\sum_{G \in \CB(\bar{\CG} \backslash \CC) \sqcup \CC} \w(G) = \sideset{}{'}\sum_{G \in \CB(\bar{\CG} \backslash \CC)} \w(G) = 1 \,.
\end{align}
From the orbit-stabilizer theorem, it follows that
\begin{align}
    \sum_{G \in \CB(\bar{\CG} \backslash \CC) \sqcup \CC} \frac{\w(G)}{\s (G)} = \frac{1}{S(\bar{\CG}_{\CC})} \,,
\end{align}
where the graph $\bar{\CG}_{\CC}$ is the graph $\bar{\CG}$ with edges in $\CC$ decorated as cuts.

The difference $S(\bar{\CG}) \neq S(\bar{\CG}_{\CC})$ is compensated for by the fact that all edges in $\CC$ must be chosen as cut propagators when applying generalized unitarity to construct the integrand. 
Apply the diagram reorganization reviewed above to Magnus diagrams, where causal propagators are treated as Feynman propagators. 
For the Magnus diagrams in the set $\CB(\bar{\CG} \backslash \CC) \sqcup \CC$, the set of cut edges $\{ e \} \supsetneq \CC$ for integrand reconstruction must include the edges in $\CC$. 
The decomposition \eqref{eq:symfac_decomp} of the symmetry factor $S(\bar{\CG}_{\CC})$ is only modified by the cut gluings $S(\{ e \})$ because of cut decorations on $\CC \subsetneq \{ e \}$, and the reorganization as the product of subamplitudes \eqref{eq:genuni_deriv} remains valid.

To construct a Magnus amplitude integrand, first construct the corresponding scattering amplitude integrand using generalized unitarity. 
Each term in the integrand has a corresponding ``Feynman diagram'' $\bar{\CG}$ that represents its propagators $1/\prod D_j$; denote this term as $\tilde{\CI} (\bar{\CG})$. 
The $i0^+$ prescription choices (causal or cut) of inverse propagators $D_j$ can be represented by ``Magnus diagrams'' $G$ that have the topology of $\bar{\CG}$. 
Define $\CF(G)[ \tilde{\CI}(\bar{\CG})]$ as the $i0^+$ prescription modification of $\tilde{\CI} (\bar{\CG})$ compatible with the graph $G$.
The $i0^+$ weight associated with $G$ is
\begin{align}
    \m (G) = \frac{\w (G)}{\s (G)} \times S (\bar{\CG}) \,. \label{eq:Murua_weight_full}
\end{align}

Let $\Omega_c (\bar\CG)$ denote the set of graphs obtained by decorating the edges of $\bar{\CG}$ with orientations or cuts (corresponding to causal or cut propagators) subject to the condition that the graph remains connected after removing all cuts. 
The corresponding Magnus amplitude is 
\begin{align}
    i M = \int \sum_{\bar{\CG}} \sum_{G \in \Omega_c (\bar{\CG})} \m (G) \CF(G) \left[ \tilde{\CI} (\bar\CG) \right] \,.
\end{align}

The argument generalizes with trivial modifications to theories that distinguish particles and antiparticles, e.g., theories with Dirac fermions, in which fermion propagators require an extra orientation. 

\subsection{Simplifying conditions for Murua \texorpdfstring{$i0^+$}{i0+} prescription}

Modern amplitude computations rely on reducing the original loop integral to a linear combination of independent \emph{master integrals} through integration-by-parts (IBP) relations. 
Under certain conditions, the Murua $i0^+$ prescription commutes with integral reduction through IBP relations, and the prescriptions can be applied at the level of master integrals. 
A sufficient condition is that all relevant underlying ``Feynman diagrams'' have unit symmetry factors, 
which is often the case when external legs are present. 

IBP relations consider a family of (scalar) loop integrals $I_{\vec{\n}}$ taking the schematic form
\begin{align}
    I_{\n_1,\cdots,\n_n} := \int_{\{ \ell \}} \frac{1}{D_1^{\n_1} \cdots D_n^{\n_n}} \,,
\end{align}
where $\vec{\n} = (\n_1, \cdots ,\n_n)$ and $\n_j \in \IZ$. 
Cut propagators and their derivatives are handled through the distributional identity $2 \pi i \, \delta(x) = (x - i0^+)^{-1} - (x + i0^+)^{-1}$.

An integral $I_{\vec{\n}}$ can have a ``Feynman diagram'' interpretation where inverse propagators $D_j$ with $\n_j > 0$ are its edges. 
Define such integrals as \emph{local} and denote the corresponding graph representation as $\bar{\CG}_{\vec{\n}}$. 
Together with external legs, nonzero powers $\n_j \neq 0$ decorate their corresponding vertices or edges. 

The $i0^+$ prescription of the integral $I_{\vec{\n}}$ is represented by the ``Magnus diagram'' $G_{\vec{\n}}$ obtained from $\bar{\CG}_{\vec{\n}}$ by further decorating the edges with orientations or cuts. 
For a local integral $I_{\vec{\n}}$ with $S(\bar{\CG}_{\vec{\n}}) = 1$, its $i0^+$ weights \label{eq:Murua_weight_full} simplify to Murua coefficients $\m (G_{\vec{\n}}) = \w (G_{\vec{\n}})$. 

IBP relations implement the transformation $\vec{\n} \to \vec{\n}'$ and are independent of propagators' $i0^+$ prescriptions. 
Assume $S(\bar{\CG}_{\vec{\n}}) = S(\bar{\CG}_{\vec{\n}'}) = 1$. 
If $\bar{\CG}_{\vec{\n}}$ and $\bar{\CG}_{\vec{\n}'}$ have the same topology, then
\begin{align}
    \m (G_{\vec{\n}'}) = \w (G_{\vec{\n}'}) = \w (G_{\vec{\n}}) = \m (G_{\vec{\n}}) \,. \label{eq:weight_cons_1}
\end{align}
If their topologies are different, some edges of $\bar{\CG}_{\vec{\n}}$ have been pinched to vertices: let $P \subset \bar{\CG}_{\vec{\n}}$ denote the set of pinched edges and $G_{\prec P}$ denote the graph obtained from $G$ by pinching the edges in $P$. 
The weight $\m (G_{\vec{\n}'})$ is equal to the sum of $\m (G_{\vec{\n}})$ over graphs $G_{\vec{\n}}$ that are isomorphic to $G_{\vec{\n}'}$ under the pinch $(G_{\vec{\n}})_{\prec P} \cong G_{\vec{\n}'}$, where the pinched edges $\ve \in P$ are directed (see \eqref{eq:Murua_cont_rule} of \textsc{End Matter}),\footnote{A cut propagator cannot be pinched during integral reduction.}
\begin{align}
    \m (G_{\vec{\n}'}) = \sum_{\substack{G_{\vec{\n}} \in \Omega_c (\bar{\CG}_{\vec{\n}}) \\ (G_{\vec{\n}})_{\prec P} \cong G_{\vec{\n}'} \\ \forall \ve \in P ,\, \ve \text{ directed} }} \m (G_{\vec{\n}}) \,. \label{eq:weight_cons_2}
\end{align}
$\Omega_c (\bar{\CG}_{\vec{\n}})$ is the set of graphs obtained by decorating the edges of $\bar{\CG}_{\vec{\n}}$ with orientations or cuts, subject to the condition that the graph remains connected after removing all cuts. 
The relations \eqref{eq:weight_cons_1} and \eqref{eq:weight_cons_2} establish the commutativity of IBP reduction and the Murua $i0^+$ prescription.
Moreover, \eqref{eq:weight_cons_2} implies that causal propagators whose $i0^+$ prescriptions do not change the value of the integral can be pinched when computing the Murua $i0^+$ prescription weights $\m (G) = \w (G)$.

\subsection{Generalized unitarity for Magnus amplitudes}

The cut-constructible parts of Magnus amplitudes can be obtained through the following procedure.
\begin{enumerate}
    \item Construct the integrand for the \emph{scattering amplitude} with the same choice of external legs and number of loops using generalized unitarity.
    \item Determine the relevant integral families $\{ I_{\vec{\n}} \}$ and the relevant local integrals (integrals with a ``Feynman diagram'' representation $\bar{\CG}_{\vec{\n}}$). If the symmetry factors of all relevant integrals are unity ($S(\bar{\CG}_{\vec{\n}}) = 1$), then reduce to master integrals. Otherwise, proceed to $i0^+$ prescription modification. \label{it:step}
    \item Modify the $i0^+$ prescription of each (master) integral to convert the scattering amplitude into the corresponding Magnus amplitude. The integral $I_{\vec{\n}} = \int \CI_{\vec{\n}}$ with graph representation $\bar{\CG}_{\vec{\n}}$ is replaced by
    \begin{align}
    \begin{gathered}
        \int \sum_{G \in \Omega_c (\bar{\CG}_{\vec{\n}})} \m (G) \CF(G) \left[ \CI_{\vec{\n}} \right] \,,
        \\ \m(G) = \frac{\w(G)}{\s(G)} \times S(\bar{\CG}_{\vec{\n}}) \,.
    \end{gathered}
    \end{align}
    Keep only the relevant integrals if the full expression is not necessary. 
    \item If integral reduction has not been performed in step \ref{it:step}, perform integral reduction.
\end{enumerate}

\section{Example: scalar--spinor QED}
Consider scalar--spinor QED in which the Pauli term introduces $\D g := (g-2) \neq 0$ at the classical level, 
\begin{align}
\begin{aligned}
    \CL &= - \frac{1}{4} F_{\m\n} F^{\m\n} + |D_\m \Phi|^2 - m_1^2 |\Phi|^2
    \\ &\phantom{=a}+ \bar{\Psi} (i \gamma^\m D_\m - m_2) \Psi + \frac{i q_2 \D g}{8 m_2} \bar{\Psi} F_{\m\n} \g^\m \g^\n \Psi \,,
\end{aligned}
\end{align}
where $D_\m \Phi = (\partial_\m - i q_1 A_\m) \Phi$ and $D_\m \Psi= (\partial_\m - i q_2 A_\m ) \Psi$. 

This theory provides an example of Magnus amplitudes with nontrivial tensor structures for which an independent computation based on worldline quantum field theory (WQFT) is available to cross-validate the results in the classical limit~\cite{Mogull:2020sak,Jakobsen:2021lvp,Kim:2023drc,Kim:2024grz}. 
Since cross-validation is the goal, only the two-photon exchange contributions in the classical limit will be considered.

\subsection{Integral coefficients}
The relevant integral coefficients correspond to the topologies
\begin{align}
\begin{aligned}
    c_{\square} &\propto \;\;\;\;\;
    \begin{fmffile}{box_coeff}
    \parbox{50pt}{
        \begin{fmfgraph*}(50,50)
            \fmfstraight
            \fmfleft{l3,l2,l1}
            \fmfright{r3,r2,r1}
            \fmf{phantom}{l2,e2,cl2,e1,cr2,e3,r2}
             \fmffreeze
            \fmf{phantom}{l1,cl1,cl2,cl3,l3}
            \fmf{phantom}{r1,cr1,cr2,cr3,r3}
             \fmffreeze
            \fmf{fermion,width=2,label=$p_2$}{r3,cr3}
            \fmf{fermion,width=2}{cr3,cr1}
            \fmf{fermion,width=2,label=$p_2'$}{cr1,r1}
            \fmf{plain,width=2,label=$p_1'$}{l1,cl1}
            \fmf{plain,width=2}{cl1,cl3}
            \fmf{plain,width=2,label=$p_1$}{cl3,l3}
            \fmf{wiggly}{cl1,cr1}
            \fmf{wiggly}{cl3,cr3}
            \fmfv{decor.shape=circle,decor.size=2}{cl1}
            \fmfv{decor.shape=circle,decor.size=2}{cl3}
            \fmfv{decor.shape=circle,decor.size=2}{cr1}
            \fmfv{decor.shape=circle,decor.size=2}{cr3}
        \end{fmfgraph*}
    }
    \end{fmffile} \;\;\;\;\;,&& & c_{\triangleright \hskip-1.5pt \triangleleft} &\propto \;\;
    \begin{fmffile}{cbox_coeff}
    \parbox{50pt}{
        \begin{fmfgraph*}(50,50)
            \fmfstraight
            \fmfleft{l3,l2,l1}
            \fmfright{r3,r2,r1}
            \fmf{phantom}{l2,e2,cl2,e1,cr2,e3,r2}
             \fmffreeze
            \fmf{phantom}{l1,cl1,cl2,cl3,l3}
            \fmf{phantom}{r1,cr1,cr2,cr3,r3}
             \fmffreeze
            \fmf{fermion,width=2}{r3,cr3,cr1,r1}
            \fmf{plain,width=2}{l1,cl1,cl3,l3}
            \fmf{wiggly}{cl1,cr3}
            \fmf{wiggly,rubout=4}{cl3,cr1}
            \fmfv{decor.shape=circle,decor.size=2}{cl1}
            \fmfv{decor.shape=circle,decor.size=2}{cl3}
            \fmfv{decor.shape=circle,decor.size=2}{cr1}
            \fmfv{decor.shape=circle,decor.size=2}{cr3}
        \end{fmfgraph*}
    }
    \end{fmffile} \;\;,
    \\ c_{\triangleleft} &\propto \;\;
    \begin{fmffile}{tri1_coeff}
    \parbox{50pt}{
        \begin{fmfgraph*}(50,50)
            \fmfstraight
            \fmfleft{l3,l2,l1}
            \fmfright{r3,r2,r1}
            \fmf{phantom}{r2,e2,cr2,e1,cl1,l2}
             \fmffreeze
            \fmf{phantom}{r1,cr1,cr2,cr3,r3}
             \fmffreeze
            \fmf{fermion,width=2}{r3,cr3,cr1,r1}
            \fmf{plain,width=2}{l3,cl1,l1}
            \fmf{wiggly}{cr3,cl1}
            \fmf{wiggly}{cr1,cl1}
            \fmfv{decor.shape=circle,decor.size=3}{cl1}
            \fmfv{decor.shape=circle,decor.size=2}{cr1}
            \fmfv{decor.shape=circle,decor.size=2}{cr3}
        \end{fmfgraph*}
    }
    \end{fmffile} \;\;,&& & c_{\triangleright} &\propto \;\;
    \begin{fmffile}{tri2_coeff}
    \parbox{50pt}{
        \begin{fmfgraph*}(50,50)
            \fmfstraight
            \fmfleft{l3,l2,l1}
            \fmfright{r3,r2,r1}
            \fmf{phantom}{l2,e2,cl2,e1,cr1,e3,r2}
             \fmffreeze
            \fmf{phantom}{l1,cl1,cl2,cl3,l3}
             \fmffreeze
            \fmf{plain,width=2}{l1,cl1,cl3,l3}
            \fmf{fermion,width=2}{r3,cr1,r1}
            \fmf{wiggly}{cl3,cr1}
            \fmf{wiggly}{cr1,cl1}
            \fmfv{decor.shape=circle,decor.size=3}{cr1}
            \fmfv{decor.shape=circle,decor.size=2}{cl1}
            \fmfv{decor.shape=circle,decor.size=2}{cl3}
        \end{fmfgraph*}
    }
    \end{fmffile} \;\;,
\end{aligned}
\end{align}
where the solid line denotes the massive scalar of mass $m_1$ with incoming (outgoing) momentum $p_1$ ($p_1'$), the solid directed line denotes the massive fermion of mass $m_2$ with incoming (outgoing) momentum $p_2$ ($p_2'$), and the wiggly lines denote the exchanged photons. 
Because all relevant underlying graphs have $S(\bar{\CG}) = 1$, the weights can be computed using $\w (G)$, and integral reduction and the Murua $i0^+$ prescription commute.

The form-factor basis is
\begin{align}
\begin{aligned}
   F_1 := \bar{u} (p_2') u (p_2) \,&,\quad F_2^{\m\n} := \bar{u} (p_2') \S^{\m\n} u (p_2) \,,
   \\ p_2' = p_2 + q \,&,\quad \g = \frac{p_1 \cdot p_2}{m_1 m_2} \,,
\end{aligned}
\end{align}
where $\S^{\m\n} = \frac{i}{4} [\g^\m , \g^\n]$ is the spin generator for the fermion. In the classical limit, $F_1 \to 2m_2$ and $F_2^{\m\n} \to 2 m_2 S_2^{\m\n}$. The latter can be recast using the spin vector $S_2^\m$ as $m_2 S_2^{\m\n} = - \e^{\m\n\a\b} p_{2\a} S_{2\b}$. 
Adopt the conventions ($D_j = - k_j^2 + m_j^2$)
\begin{align}
    i A &= \sum_\# c_\# I_\# \,,\quad I = \int \frac{d^D \ell}{(2\pi)^D} \prod_j \frac{1}{D_j} \,.
\end{align}
The leading soft-expansion ($t = q^2$ expansion) terms---the terms relevant for the classical limit---of the integral coefficients are ($\e[a,b,c,d]:=\e_{\m\n\l\s} a^\m b^\n c^\l d^\s$ with $\e^{0123}=+1$)
\begin{align}
    c_{\square} &= 4 q_1^2 q_2^2 \left[ 2 m_1^2 m_2 \g^2 \, F_1 + i m_1 (2+ \D g) \g \, p_{1\m} q_\n F_2^{\m\n} \right] \nn
    \\ &\to 8 q_1^2 q_2^2 \left[ 2 m_1^2 m_2^2 \g^2 + i m_1 (2+ \D g) \g \, \e [ q, p_{1}, p_{2}, S_{2}] \right] \,,
    \\ c_{\triangleright \hskip-1.5pt \triangleleft} &= 4 q_1^2 q_2^2 \left[ 2 m_1^2 m_2 \g^2 \, F_1 + i m_1 (2+ \D g) \g \, p_{1\m} q_\n F_2^{\m\n} \right] \nn
    \\ &\to 8 q_1^2 q_2^2 \left[ 2 m_1^2 m_2^2 \g^2 + i m_1 (2+ \D g) \g \, \e [ q, p_{1}, p_{2}, S_{2}] \right] \,,
    \\ c_{\triangleleft} &= 4 q_1^2 q_2^2 \left[ m_2 \, F_1 + i \frac{ (2+\D g)\g }{m_1 (\g^2 - 1)} \, p_{1\m} q_\n F_2^{\m\n} \right] \nn
    \\ &\to 8 q_1^2 q_2^2 \left[ m_2^2 + i \frac{ (2+\D g)\g }{m_1 (\g^2 - 1)} \, \e [ q, p_{1}, p_{2}, S_{2}] \right] \,,
    \\ c_{\triangleright} &= 4 q_1^2 q_2^2 \left[ \frac{m_1^2}{m_2} \, F_1 + i \frac{ m_1 (1+\D g)\g }{m_2^2 (\g^2 - 1)} \, p_{1\m} q_\n F_2^{\m\n} \right] \nn
    \\ &\to 8 q_1^2 q_2^2 \left[ m_1^2 + i \frac{ m_1 (1+\D g)\g }{m_2^2 (\g^2 - 1)} \, \e [ q, p_{1}, p_{2}, S_{2}] \right] \,.
\end{align}

\subsection{Classical limit in impact parameter space}

The one-loop master integrals fall into two classes: (1) integrals consisting solely of causal propagators, and (2) integrals with one cut propagator. 
The first class can be discarded because only the vanishing integrals (integrals with closed causal loops) have nonvanishing weights---see Table~\ref{tb:Murua_one-loop}. For the second class of integrals, only the integrals with cuts imposed on the massive propagators are relevant for the classical limit.

\begin{table*}[t]
    \caption{\label{tb:Murua_one-loop}%
        Murua coefficients for directed one-loop topologies. Only circular causal prescriptions have nonvanishing weights; therefore, all one-loop integrals consisting solely of causal propagators either have vanishing weights or vanishing values.
    }
    \begin{ruledtabular}
        \begin{tabular}{ccccccccccc}
            \rule[-17pt]{0pt}{36pt}%
            $G$ & &
    \begin{fmffile}{bubble1}
    \parbox{25pt}{
        \begin{fmfgraph*}(25,25)
            \fmfstraight
            \fmfleft{b1}
            \fmfright{t1}
            \fmf{fermion,left=0.5}{t1,b1}
            \fmf{fermion,right=0.5}{t1,b1}
            \fmfv{decor.shape=circle,decor.size=2}{b1}
            \fmfv{decor.shape=circle,decor.size=2}{t1}
        \end{fmfgraph*}
    }
    \end{fmffile}
            & 
    \begin{fmffile}{bubble2}
    \parbox{25pt}{
        \begin{fmfgraph*}(25,25)
            \fmfstraight
            \fmfleft{t1}
            \fmfright{b1}
            \fmf{fermion,right=0.5}{t1,b1}
            \fmf{fermion,right=0.5}{b1,t1}
            \fmfv{decor.shape=circle,decor.size=2}{b1}
            \fmfv{decor.shape=circle,decor.size=2}{t1}
        \end{fmfgraph*}
    }
    \end{fmffile}
            & & 
    \begin{fmffile}{triangle1}
    \parbox{25pt}{
        \begin{fmfgraph*}(25,25)
            \fmfstraight
            \fmfleft{l1,l2}
            \fmfright{r1,r2}
            \fmftop{t1}
            \fmf{fermion}{l1,t1}
            \fmf{fermion}{l1,r1,t1}
            \fmfv{decor.shape=circle,decor.size=2}{t1}
            \fmfv{decor.shape=circle,decor.size=2}{l1}
            \fmfv{decor.shape=circle,decor.size=2}{r1}
        \end{fmfgraph*}
    }
    \end{fmffile}
            &
    \begin{fmffile}{triangle2}
    \parbox{25pt}{
        \begin{fmfgraph*}(25,25)
            \fmfstraight
            \fmfleft{l1,l2}
            \fmfright{r1,r2}
            \fmftop{t1}
            \fmf{fermion}{l1,r1,t1,l1}
            \fmfv{decor.shape=circle,decor.size=2}{t1}
            \fmfv{decor.shape=circle,decor.size=2}{l1}
            \fmfv{decor.shape=circle,decor.size=2}{r1}
        \end{fmfgraph*}
    }
    \end{fmffile}
            & & 
    \begin{fmffile}{box1}
    \parbox{25pt}{
        \begin{fmfgraph*}(25,25)
            \fmfstraight
            \fmfleft{l1,l2}
            \fmfright{r1,r2}
            \fmf{fermion}{l1,l2}
            \fmf{fermion}{l1,r1,r2,l2}
            \fmfv{decor.shape=circle,decor.size=2}{l1}
            \fmfv{decor.shape=circle,decor.size=2}{l2}
            \fmfv{decor.shape=circle,decor.size=2}{r1}
            \fmfv{decor.shape=circle,decor.size=2}{r2}
        \end{fmfgraph*}
    }
    \end{fmffile}
            & 
    \begin{fmffile}{box2}
    \parbox{25pt}{
        \begin{fmfgraph*}(25,25)
            \fmfstraight
            \fmfleft{l1,l2}
            \fmfright{r1,r2}
            \fmf{fermion}{r1,l1,l2}
            \fmf{fermion}{r1,r2,l2}
            \fmfv{decor.shape=circle,decor.size=2}{l1}
            \fmfv{decor.shape=circle,decor.size=2}{l2}
            \fmfv{decor.shape=circle,decor.size=2}{r1}
            \fmfv{decor.shape=circle,decor.size=2}{r2}
        \end{fmfgraph*}
    }
    \end{fmffile}
            & 
    \begin{fmffile}{box3}
    \parbox{25pt}{
        \begin{fmfgraph*}(25,25)
            \fmfstraight
            \fmfleft{l1,l2}
            \fmfright{r1,r2}
            \fmf{fermion}{l1,r1,r2,l2,l1}
            \fmfv{decor.shape=circle,decor.size=2}{l1}
            \fmfv{decor.shape=circle,decor.size=2}{l2}
            \fmfv{decor.shape=circle,decor.size=2}{r1}
            \fmfv{decor.shape=circle,decor.size=2}{r2}
        \end{fmfgraph*}
    }
    \end{fmffile}
    \\
            $\w(G)$ & & $0$ & $\frac{1}{2}$ & & $0$ &
            $\frac{1}{2}$ & & $0$ & $0$ & $\frac{1}{2}$ \\
        \end{tabular}
    \end{ruledtabular}
\end{table*}

Consider the left-cut-box integral where the massive propagator of mass $m_1$ is a cut propagator. Because of the on-shell conditions imposed by the external legs and the cut propagator, the massless propagators cannot go on shell and their $i0^+$ prescriptions are irrelevant. 
The Murua prescription for the left-cut-box integral is obtained by contracting the massless propagators to vertices and becomes the ``average'' of $i0^+$ prescriptions,
\begin{align}
\begin{aligned}
    I_{\raisebox{0.2ex}{$\scriptstyle-$}\!\!\!\square} &= \w \! \left( 
    \begin{fmffile}{left_cut_box1}
    \parbox{10pt}{
        \begin{fmfgraph*}(10,25)
            \fmfkeep{lcb1}
            \fmfstraight
            \fmftop{t1}
            \fmfbottom{b1}
            \fmf{fermion,right=0.32}{b1,t1}
            \fmf{dashes,left=0.32}{b1,t1}
            \fmfv{decor.shape=circle,decor.size=2}{b1}
            \fmfv{decor.shape=circle,decor.size=2}{t1}
        \end{fmfgraph*}
    }
    \end{fmffile}
    \right) I_{\raisebox{0.2ex}{$\scriptstyle-$}\!\!\!\square}^{+} + \w \! \left( 
    \begin{fmffile}{left_cut_box2}
    \parbox{10pt}{
        \begin{fmfgraph*}(10,25)
            \fmfkeep{lcb2}
            \fmfstraight
            \fmftop{t1}
            \fmfbottom{b1}
            \fmf{fermion,left=0.32}{t1,b1}
            \fmf{dashes,left=0.32}{b1,t1}
            \fmfv{decor.shape=circle,decor.size=2}{b1}
            \fmfv{decor.shape=circle,decor.size=2}{t1}
        \end{fmfgraph*}
    }
    \end{fmffile}
    \right) I_{\raisebox{0.2ex}{$\scriptstyle-$}\!\!\!\square}^{-}
    = \frac{I_{\raisebox{0.2ex}{$\scriptstyle-$}\!\!\!\square}^{+} + I_{\raisebox{0.2ex}{$\scriptstyle-$}\!\!\!\square}^{-} }{2} \simeq 0 \,,\!
\end{aligned}
\end{align}
which vanishes. 
The same analysis applies to the right-cut-box integral and the cut-crossed-box integrals; therefore, the box ($c_{\square}$) and the crossed-box ($c_{\triangleright \hskip-1.5pt \triangleleft}$) integral coefficients can be discarded in the classical limit.

The same reasoning also applies to the left-cut-triangle integral $I_{-\!\!\!\triangleright}$, so the $i0^+$ prescriptions of its massless propagators are irrelevant. The integral $I_{-\!\!\!\triangleright}$ is
\begin{align}
    I_{-\!\!\!\triangleright} &\simeq \frac{i}{4 m_1} \frac{\G (\frac{5-D}{2}) \G(\frac{D-3}{2})^2}{(4 \pi)^{\frac{D-1}{2}} \G(D-3)} \frac{1}{[\vec{q}^2]^{\frac{5-D}{2}}} \,,
\end{align}
and the right-cut-triangle integral $I_{\triangleleft\!\!\!-}$ is obtained by exchanging $m_1 \leftrightarrow m_2$.

In $D=4$, Fourier-transforming the Magnus amplitude to impact parameter space yields
\begin{align}
\begin{aligned}
    i \chi &= \int \frac{d^D q}{(2 \pi)^D} \; 2 \pi \delta [2 (p_1 \cdot q)] \; 2 \pi \delta [2 (p_2 \cdot q)]
    \\ &\phantom{=asdfasdfasdf} \times e^{i b \cdot q} \left( c_{\triangleright} I_{-\!\!\!\triangleright} + c_{\triangleleft} I_{\triangleleft\!\!\!-} \right)
    \\ &= i \; \frac{q_1^2 q_2^2}{32\pi m_1 m_2} \Bigg[ \frac{(m_1 + m_2)}{\sqrt{\g^2 - 1} |\vec{b}|} 
    \\ &\phantom{=} + \frac{\g \left( (2 + \D g) m_2 + (1 + \D g) m_1 \right) \e[b, p_1, p_2 , S_2]}{m_1 m_2^2 (\g^2 - 1)^{3/2} |\vec{b}|^3} \Bigg] \,.
\end{aligned} \label{eq:FT_Magnusian}
\end{align}
This Fourier transform can be viewed as computing the expectation value of $\chi$ in a wavepacket state~\cite{Kosower:2018adc,Kim:2025gis,Jeong:2026pyw}.
This result is consistent with the WQFT-based computation of the Magnusian---see \textsc{End Matter}.

\section{Discussion}

It was shown how generalized unitarity for scattering amplitudes can be applied to compute Magnus amplitudes based on the mathematical properties of the Murua coefficients $\w (G)$. 
The key observation was that Magnus amplitude integrands can be obtained from scattering amplitude integrands by modifying the $i0^+$ prescriptions of the propagators. 
A procedure for adapting unitarity methods to Magnus amplitudes was proposed and applied to the computation of the one-loop scalar--spinor QED Magnus amplitude in the classical limit.
The result has been checked against the corresponding WQFT computations, and the consistency of the two computations supports the validity of the proposal.

One future goal is to understand the precise conditions under which integral reduction and the Murua $i0^+$ prescription commute. 
Although a sufficient condition has been identified and exploited in the presented example, it seems possible to relax the conditions for commutativity; e.g., commutativity appears to hold when all vertices can be distinguished. 
Also, understanding how the symmetries of integral families interact with the Murua $i0^+$ prescription would benefit large-scale computations.

The modern amplitudes philosophy can be summarized as ``Feynman diagrams are not the most efficient way of computing scattering amplitudes.'' 
This work brings Magnus amplitudes closer to this philosophy: Magnus diagrams are not the most efficient way of computing Magnus amplitudes, and the tools developed for scattering amplitudes can also be applied to Magnus amplitudes---with some modifications. 
A next step would be to determine whether other methods developed for scattering amplitudes could also be adapted to Magnus amplitudes.

Many modern techniques for computing scattering amplitudes rely on our understanding of the analyticity properties of scattering amplitudes~\cite{Goncharov:2010jf,Dixon:2011pw,Dixon:2013eka,Golden:2013xva,Drummond:2014ffa,Caron-Huot:2016owq}. 
Therefore, another immediate goal would be to develop tools for understanding the analytic structures of Magnus amplitudes and the Magnusian---the logarithm of the $S$-matrix. 
One may remark that this amounts to the shift of focus from the ``unitarity of the analytic $S$-matrix'' to the ``analyticity of the unitary $S$-matrix.''


\textbf{Acknowledgments} --- 
The authors thank Li Guo, Joon-Hwi Kim, and Sangmin Lee for initial collaboration and insightful discussions. 
JWK appreciates fruitful discussions with Yu-Tin Huang, Gabriele Travaglini, Christopher White, and Alexander Zhiboedov.
\texttt{Mathematica} packages \texttt{xAct}~\cite{xAct} and \texttt{LiteRed}~\cite{Lee:2012cn} were used in some of the computations for this work.
AI (GPT-6 Pro) was used to compile references, aid formatting, and perform independent computations for comparison with the presented results. 
SK was supported
by the National Research Foundation of Korea (NRF) grant NRF RS-2024-00351197 and KIAS
grant PG006002. JL was supported by the Beatriz Galindo Senior grant BG24/00114 from the Spanish Ministry of Science, Innovation and Universities.


\bibliography{manuscript.bib,noinspire.bib}

%

\clearpage
\onecolumngrid
\section*{End Matter}
\subsection{Murua coefficient normalization}
The Murua coefficient $\w (G)$ is related to the Murua coefficient $\w_c (G)$ of Ref.~\cite{Guo:2026hyi} (with $(\b,\g) = (0,1)$), $\w_{\text{bw}}(G)$ of Ref.~\cite{Guo:2026xaw}, and $\w_{\text{BB+}}(G)$ of Ref.~\cite{Brandhuber:2025igz} by the relation ($\w_{\text{bw}} (G)$ and $\w_{\text{BB+}}(G)$ are only defined for acyclic graphs)
\begin{align}
    \w (G) = (-1)^{E_d(G)} \w_c (G) \stackrel{\cdot}{=} (-1)^{\frac{V(G) + E_d (G) - 1}{2}} \w_{\text{bw}}(G) = 2^{|\CC (G)|} \w_{\text{BB+}}(G)\,,
\end{align}
where $E_d(G)$ is the number of directed edges, $V(G)$ is the number of vertices, $|\CC (G)|$ is the number of cut propagators, and $\stackrel{\cdot}{=}$ denotes equality when both sides are defined.
The Murua coefficients that contribute to Magnus amplitudes are all real since $L_d(G) = E_d (G) - V(G) + 1$ is the number of loops that consist solely of directed edges and $\w_{\text{bw}}(G)$ is only relevant for even $L_d(G)$.
In this normalization the Murua coefficients satisfy the relations
\begin{align}
    \w (G) + \w (G_{\bar{\ve}}) = \w (G_{\prec \ve}) \,,\quad \w(G) = \w (G\backslash c) \,, \label{eq:Murua_cont_rule}
\end{align}
where $G_{\bar{\ve}}$ is $G$ with the directed edge $\ve$ reversed, $G_{\prec \ve}$ is $G$ with the directed edge $\ve$ contracted to a vertex, and $G \backslash c$ is $G$ with a cut propagator $c \in \CC (G)$ removed (if $G \backslash c$ becomes disconnected, $\w(G) = 0$).

\subsection{Normalization conventions for Feynman rules}
The matrix elements are normalized as
\begin{align}
    i (2\pi)^D \delta^D (P_F^\m - P_I^\m) \,  A_{I \to F} = \bra{F} S - \iden \ket{I} \,,\quad i (2\pi)^D \delta^D (P_F^\m - P_I^\m) \,  M_{I \to F} = \bra{F} \log S \ket{I} \,.
\end{align}
The valence-$n$ vertex rules $\CV_n$ are derived from the interaction action $S_{\text{int}} = - \int H_{\text{int}}$ as
\begin{align}
    \CV_n = i \frac{\delta^n S_{\text{int}}}{\delta \phi^n} = - i \frac{\delta^n}{\delta \phi^n} \int H_{\text{int}} \,.
\end{align}
Adopting the positive frequency expansion convention $\phi (x) = \int_p \phi (p) \, e^{- i p x}$ results in the all-incoming convention for the vertex rules. 
The Feynman/retarded propagators are given as
\begin{align}
    \la \phi(p) \phi(-p) \ra = \int_x e^{i p x} \la \phi (x) \phi (0) \ra = \frac{- i}{- p^2 + m^2 - i \eta 0^+} \,,\quad 
    \eta = \left\{
    \begin{aligned}
        &1 && & \text{Feynman}
        \\ & p^0 && & \text{retarded}
    \end{aligned}
    \right. \,,
\end{align}
where modifiers (time-ordering operator/time-ordering step function and commutator) have been suppressed for brevity. The Hadamard cut propagator is normalized as
\begin{align}
    \frac{1}{2}\la \{ \phi(p) , \phi (-p) \} \ra = \frac{1}{2} \int_x e^{i p x} \la \phi (x) \phi (0) + \phi (0) \phi(x) \ra = \pi \delta (- p^2 + m^2) \,,
\end{align}
which implements the substitution $(-p^2 + m^2)^{-1} \to i \pi \delta (- p^2 + m^2)$ for cut propagators.

The Hermiticity of the Magnusian $\chi$ demands the following: (1) diagrams consisting solely of directed edges either have vanishing weights or vanishing integrals at odd loop orders, (2) nonvanishing contributions at odd loop orders must have an odd number of cut propagators, (3) nonvanishing contributions at even loop orders must have an even number of cut propagators~\cite{Brandhuber:2025igz,Guo:2026xaw}; e.g., see Table~\ref{tb:Murua_one-loop}.

\subsection{Master integrals with one cut propagator}
Define
\begin{align}
    I_{\raisebox{0.2ex}{$\scriptstyle-$}\!\!\!\square}^{\pm} &= \int \frac{d^D \ell}{(2\pi)^D} \frac{i \pi \delta [-(\ell - p_1)^2 + m_1^2]}{[- \ell^2] [- ( \ell^0 + p_2^0 \pm i 0^+)^2 + (\vec{\ell} + \vec{p}_2)^2 + m_2^2] [ - (\ell - q)^2 ]} \,.
\end{align}
Due to the Fourier transform kernel $\propto \delta [2 (p_1 \cdot q)] \times \delta [2 (p_2 \cdot q)]$, the integral can be evaluated under the constraints $p_1 \cdot q = p_2 \cdot q = 0$.\footnote{Although these constraints are inconsistent with the (real) on-shell kinematics, this na\"ive treatment does not affect the classical limit at the considered order in perturbation theory. A more rigorous treatment of the $\hbar$ expansion is available through the heavy-mass effective field theory/heavy-particle effective theory framework~\cite{Damgaard:2019lfh,Brandhuber:2021eyq}, which reveals an evanescent term as $D \to 4$.} 
In the rest frame of $p_1$, the cut propagator can be approximated as ($E_\ell = \sqrt{m_1^2 + \vec{\ell}^2}$)
\begin{align}
    \delta [-(\ell - p_1)^2 + m_1^2] &= \frac{\delta ( \ell^0 - m_1 + E_\ell ) + \delta (\ell^0 - m_1 - E_\ell)}{2 E_\ell} \simeq \frac{\delta (\ell^0)}{2 m_1} \,,
\end{align}
at leading order in the soft ($\frac{\sqrt{-q^2}}{m}$) expansion, which yields (see Appendix B of Ref.~\cite{Kim:2024grz})
\begin{align}
    I_{\raisebox{0.2ex}{$\scriptstyle-$}\!\!\!\square}^{\pm} &\simeq \frac{i}{4 m_1} \int \frac{d^{D-1} \ell_E}{(2\pi)^{D-1}} \frac{1}{\vec{\ell}^2(\vec{\ell} - \vec{q})^2 (2 \vec{\ell} \cdot \vec{p}_2 \mp i p_2^0 0^+)} = \frac{\mp \pi}{4 m_1 m_2 \sqrt{\g^2 - 1}} \frac{\G (\frac{D-4}{2})^2 \G( \frac{6-D}{2})}{ (4 \pi)^{\frac{D}{2}} \G (D-4) } \frac{1}{(\vec{q}^2)^{\frac{6-D}{2}}} \,,
    \\ \therefore I_{\raisebox{0.2ex}{$\scriptstyle-$}\!\!\!\square} &= \frac{I_{\raisebox{0.2ex}{$\scriptstyle-$}\!\!\!\square}^+ + I_{\raisebox{0.2ex}{$\scriptstyle-$}\!\!\!\square}^-}{2} \simeq 0 \,,
\end{align}
where $d^{D-1} \ell_E$ is the Euclidean loop momentum measure. 
The same analysis applies to the left-cut-triangle integral $I_{-\!\!\!\triangleright}$ defined as
\begin{align}
    I_{-\!\!\!\triangleright} &= \int \frac{d^D \ell}{(2\pi)^D} \frac{i \pi \delta [-(\ell - p_1)^2 + m_1^2]}{[- \ell^2] [ - (\ell - q)^2 ]} \simeq \frac{i}{4 m_1} \int \frac{d^{D-1} \ell_E}{(2\pi)^{D-1}} \frac{1}{\vec{\ell}^2(\vec{\ell} - \vec{q})^2} = \frac{i}{4 m_1} \frac{\G (\frac{5-D}{2}) \G(\frac{D-3}{2})^2}{(4 \pi)^{\frac{D-1}{2}} \G(D-3)} \frac{1}{[\vec{q}^2]^{\frac{5-D}{2}}} \,,
\end{align}
where $\delta [-(\ell - p_1)^2 + m_1^2] \simeq \delta[2 (\ell \cdot p_1)] = \delta (2 \ell^0 m_1)$ has been used to evaluate the classical limit of the integral in the rest frame of $p_1$. The right-cut-triangle integral $I_{\triangleleft\!\!\!-}$ is obtained by exchanging $m_1 \leftrightarrow m_2$.

\subsection{WQFT comparison computation}
A comparison computation is available through the WQFT setup of Ref.~\cite{Kim:2023drc}. The supersymmetric worldline action for a spinning massive particle of charge $q$ and gyromagnetic ratio $g = 2 + \D g$ coupled to the Maxwell field $A_\m$ is given as
\begin{align}
    S_{\text{WL}} &= - \int d \t \left[ \frac{m}{2} \dot{x}^2 + q A \cdot \dot{x} + \frac{i}{2} m \psi \cdot \dot{\psi} - \frac{i (2 + \D g)q}{4} F_{\m\n} \psi^\m \psi^\n + \frac{i \D g q (\dot{x} \cdot \psi) F_{\m\n} \dot{x}^\m \psi^\n}{2 \dot{x}^2} \right] \,,
\end{align}
where the spin supplementary condition (SSC) parameter is set to $h = 1$ for SSC preservation. 
The details of the WQFT setup and the derivation of Feynman rules can be found in Ref.~\cite{Kim:2023drc}. 
The resulting one-loop Magnusian is
\begin{align}
    \chi &= \frac{(m_1 + m_2)q_1^2 q_2^2}{32 \pi m_1 m_2 \sqrt{\g^2 - 1} |\vec{b}|} + \frac{\g \left( (2 + \D g) m_2 + (1 + \D g) m_1 \right) q_1^2 q_2^2 \e[b, p_1, p_2, S_2]}{32 \pi m_1^2 m_2^3 (\g^2 - 1)^{3/2} |\vec{b}|^3} \,,
\end{align}
which agrees with the scalar--spinor QED computation in the main text \eqref{eq:FT_Magnusian}. Setting $\D g = 0$ reduces this result to the linear-in-spin truncation of the results presented in Ref.~\cite{Kim:2024grz} (up to a difference in the convention for $\e_{\m\n\l\s}$).

\end{document}